\documentclass[10pt,twocolumn,letterpaper]{article}
\usepackage[utf8]{inputenc}
\usepackage{amsmath}
\usepackage{graphicx}
\usepackage{hyperref}
\usepackage{graphicx} 
\usepackage{adjustbox} 
\usepackage{booktabs}
\usepackage{multirow}
\usepackage{pgfplots}
\pgfplotsset{compat=1.18}
\usepackage{amsmath}

\title{Efficiency and Effectiveness of SPLADE Models on Billion-Scale Web Document Title}
\author{Taeryun Won, Tae Kwan Lee, Hiun Kim, Hyemin Lee \\ NAVER Search \\ \texttt{\{lory.tail, taekwan.lee, hiun.kim, hmin.lee\}@navercorp.com}}

\date{}

\begin{document}

\maketitle

\begin{abstract}
This paper presents a comprehensive comparison of BM25, SPLADE, and Expanded-SPLADE models in the context of large-scale web document retrieval. We evaluate the effectiveness and efficiency of these models on datasets spanning from tens of millions to billions of web document titles. SPLADE and Expanded-SPLADE, which utilize sparse lexical representations, demonstrate superior retrieval performance compared to BM25, especially for complex queries. However, these models incur higher computational costs. We introduce pruning strategies, including document-centric pruning and top-k query term selection, boolean query with term threshold to mitigate these costs and improve the models' efficiency without significantly sacrificing retrieval performance. The results show that Expanded-SPLADE strikes the best balance between effectiveness and efficiency, particularly when handling large datasets. Our findings offer valuable insights for deploying sparse retrieval models in large-scale search engines.
\end{abstract}

\section{Introduction}

Information retrieval systems have become essential in the digital age, where users are confronted with vast amounts of data, often numbering in the billions. Traditional retrieval models such as BM25 have been widely adopted for their effectiveness and simplicity. However, these models often struggle to capture the semantic richness of queries and documents \cite{robertson2009bm25,guo2016deep,mitra2018neural}.

In recent years, sparse retrieval models like SPLADE (Sparse Lexical and Expansion Model for Information Retrieval) have emerged as a promising solution, leveraging sparse representations for more efficient and effective retrieval \cite{formal2021splade,formal2021spladev2}. SPLADE improves upon traditional retrieval methods by learning sparse lexical representations, allowing for better query expansion and document ranking without the computational overhead of dense retrieval models .

This paper presents a comparative analysis of BM25, SPLADE, and its extended variant, Expanded-SPLADE \cite{dudek2022expanded}, on a billion-scale web document dataset. We evaluate both the \textit{effectiveness}—measured through retrieval quality—and the \textit{efficiency}—considering computational resources like latency and FLOPS (floating-point operations per second). By examining these models on both small and large-scale datasets, we provide insights into the trade-offs between effectiveness and efficiency in large-scale web search applications. 

We also explore how pruning strategies, such as document-centric pruning, top-k query term selection, and boolean query with term threshold, can mitigate the computational overhead of sparse models like SPLADE and Expanded-SPLADE, making them more suitable for real-world search engines \cite{static_pruning, lassance2022efficiency, mallia2024fastpruning}.

\section{Background}

The BM25 model has long been a standard in information retrieval, known for its effectiveness in ranking documents based on term frequency and inverse document frequency \cite{robertson2009bm25}. Its probabilistic framework allows it to capture the relevance of documents to a query based on the occurrence and distribution of terms. Despite its popularity, BM25 has limitations in handling complex, semantically-rich queries due to its reliance on keyword matching.

SPLADE models address this issue by employing sparse lexical representations learned over a fixed vocabulary. SPLADE models expand the representation of both queries and documents by incorporating learned embeddings that are sparse in nature, allowing them to capture deeper semantic relationships. Furthermore, SPLADE \cite{formal2021spladev2,lassance2022efficiency} has demonstrated superior performance compared to other sparse retrieval models \cite{contextaware2021,doc2query2020,sparterm2022,coil2021,passageimpacts2021} and dense retrieval models \cite{xiong2021approximate,hofstatter2021inbatch,qu2021efficiently}, effectively balancing efficiency and retrieval effectiveness while maintaining strong semantic understanding

Traditional SPLADE models are inherently limited by the WordPiece vocabulary of BERT, which constrains their lexical representation power. To overcome this limitation, recent work by \cite{dudek2022expanded} proposes a novel approach that learns sparse lexical representations over an expanded vocabulary, allowing for greater flexibility in encoding document-query interactions. Empirical results demonstrate that Expanded-SPLADE maintains performance comparable to WordPiece-based SPLADE while offering enhanced adaptability by supporting custom output vocabularies. This flexibility makes it particularly useful for retrieval tasks that require a broader lexical coverage without sacrificing retrieval efficiency.

In the context of large-scale information retrieval, managing computational efficiency is crucial. Sparse retrieval models like SPLADE and Expanded-SPLADE, though effective, often require significant computational resources. Recent work has explored techniques such as pruning strategies to reduce latency and computational costs without significantly sacrificing retrieval performance \cite{static_pruning, lassance2022efficiency}. These strategies include document-centric static pruning \cite{buettcher2006document, wicaksono2008within}, where less important terms are eliminated from documents, and the use of a smaller PLM query encoder to improve efficiency while maintaining retrieval effectiveness.

\section{Methodology}

\subsection{Training Data}
Our SPLADE model was trained on a large-scale Korean dataset consisting of web document titles and real-world user queries. Specifically, the dataset contains 260 million (query, document) pairs, with 115 million unique documents and 28 million unique queries collected from our internet search engine. These queries reflect authentic user search behavior, enhancing the realism of our retrieval task.

\subsection{Evaluation Data}
To evaluate the effectiveness of our model, we use two distinct evaluation sets: small and large. Both evaluation datasets consist of Korean text data.

\subsubsection{Small Evaluation Set}
\begin{itemize}
    \item \textbf{Number of Documents}: 20 million web document titles
    \item \textbf{Average Document Title Length}: 13.4 words
    \item \textbf{Query Count}: 8,936 queries
    \item \textbf{Average Query Length}: 2.6 words
\end{itemize}

\subsubsection{Large Evaluation Set}
\begin{itemize}
    \item \textbf{Number of Documents}: 9 billion web document titles
    \item \textbf{Average Document Title Length}: 10.3 words
    \item \textbf{Query Count}: 1,000 queries
    \item \textbf{Average Query Length}: 4.5 words
\end{itemize}

\subsection{Training Approach}

We trained two models: SPLADE and Expanded-SPLADE. The SPLADE model utilizes a vocabulary size of 32,000, whereas the Expanded-SPLADE model expands this to 100,000 words.

For model initialization, we used an in-house Korean BERT model as the pretrained backbone \cite{devlin2019bert}.

Unlike the original SPLADE paper, both models were trained using only in-batch negative loss, which is formulated as:
\begin{equation}
    \mathcal{L}_{\text{in-batch}} = - \frac{1}{|B|} \sum_{(q,d^+)} \log \frac{e^{s(q,d^+)}}{e^{s(q,d^+)} + \sum\limits_{d^- \in B} e^{s(q,d^-)}}
\end{equation}
where $s(q,d)$ represents the similarity score between query $q$ and document $d$, $d^+$ is the relevant document, $d^-$ are negative samples drawn from the batch $B$, and $|B|$ is the batch size. This decision was made to simplify the sampling process. Additionally, the Expanded-SPLADE model underwent Masked Language Model (MLM) pretraining before fine-tuning. This pretraining phase leveraged a dataset containing 600 million web documents, from which the 100,000 most frequent words were extracted as the vocabulary.

For regularization, the SPLADE model employs the FLOPS loss, which is defined as:
\begin{equation}
    \mathcal{L}_{\text{FLOPS}}(T) = \bar{w}^{(T)} \cdot \bar{w}^{(T)}
\end{equation}
where $\bar{w}^{(T)}$ denotes the mean representation of the output embeddings across all texts in $T$, and $\cdot$ signifies the inner product operation \cite{paria2020minimizing}.

In contrast, the Expanded-SPLADE model utilizes the joint FLOPS regularization loss, which is defined as:
\begin{equation}
    \mathcal{L}_{\text{jFLOPS}}(Q, D) = \bar{w}^{(Q)} \cdot \bar{w}^{(D)}
\end{equation}

This formulation ensures that the embeddings of queries and documents are effectively regularized, aligning their representations in the embedding space for improved retrieval performance \cite{dudek2022expanded}.

\subsection{Semantic Similarity Score (SSS)}
To evaluate retrieval effectiveness, we employ the Semantic Similarity Score (SSS), which quantifies the semantic relevance between retrieved documents and ground-truth documents. Unlike keyword-based evaluation metrics, SSS leverages learned embeddings to compute cosine similarity, providing a more robust measure of semantic relevance. Given a retrieved document \( d_i \) and a ground-truth document \( g_j \), SSS is defined as:

\begin{equation}
SSS(d_i, g_j) = \text{sim}(\text{Embedding}(d_i), \text{Embedding}(g_j))
\end{equation}

where similarity is computed using cosine similarity. The final retrieval performance is computed as the mean SSS across all retrieved documents for a query. A higher SSS indicates stronger semantic relevance. In our implementation, we use a company-specific Korean sentence embedding model.

\subsection{Computational Efficiency}
To provide a deeper understanding of computational efficiency, we evaluated the models using FLOPS and latency. For sparse models, the number of operations is significantly influenced by the sparsity level, as fewer non-zero elements lead to fewer computations. In addition to FLOPS, we measure latency to evaluate the real-time efficiency of each model. Latency is defined as the total time taken to process a query and retrieve relevant documents. We report latency in seconds per query (s/query), measured using our in-house search engine in an experimental environment where it was intentionally overloaded with data, rather than in a real-world application setting.

\subsection{Pruning Strategies}
Efficient retrieval in large-scale information retrieval (IR) systems requires pruning techniques to manage computational resources effectively. In this work, we employ three pruning strategies:

\textbf{1. Document-Centric Static Pruning:} Document-centric pruning is a static index compression technique that removes less important information at the document level to enhance search efficiency. This approach reduces the index size by eliminating terms or data within individual documents that contribute minimally to retrieval quality. In our study, we employ a top-k term selection approach, where only the k most important terms are retained for each document.

\textbf{2. Top-k Query Term Selection:} Top-k Query Term Selection is a real-time query pruning technique that enhances search efficiency by selecting only the k most important terms from a given query. Instead of processing all query terms, this method retains only those with the highest impact scores. By eliminating less significant terms, this approach reduces computational overhead while maintaining retrieval effectiveness.

\textbf{3. Boolean Query with Term Threshold:} Traditional Boolean queries operate based on strict AND/OR conditions, which can either be too restrictive (AND) or too permissive (OR). To balance these extremes, we employ Boolean Query with Term Threshold, where a document is considered relevant only if it contains a minimum threshold of query terms. The application of threshold conditions in Boolean queries allows for more fine-grained control over relevance scoring, making them adaptable to real-world search applications. Additionally, threshold-based approaches in high-dimensional spaces have been shown to optimize query performance while maintaining optimality guarantees \cite{li2018cosine}. Our retrieval pipeline consists of two main steps: (1) Initial Candidate Retrieval, where we retrieve a set of candidate documents using a should-based query in the search engine, ensuring that at least one query term is present, and (2) Threshold-based Filtering, where only documents meeting a predefined threshold (e.g., 50\%, 80\%) of matching query terms are retained. The filtered results are then ranked using similarity metrics such as BM25 or SPLADE and evaluated for retrieval effectiveness. This approach helps in reducing the overall number of documents processed at later stages, leading to improvements in computational efficiency.

\section{Results}

\subsection{Effectiveness vs. Efficiency Trade-off Across Different Dataset Sizes}
\textbf{Performance on Small Dataset}: Figure 1 shows the effectiveness vs. efficiency trade-off for BM25, SPLADE, and Expanded-SPLADE on the small dataset. Effectiveness is measured using MRR@10, while efficiency is assessed in terms of FLOPS. SPLADE outperforms BM25 in terms of effectiveness but at the cost of increased FLOPS. Expanded-SPLADE achieves a balance, showing significant improvement over BM25 while keeping computational cost lower than SPLADE.

\textbf{Performance on Large Dataset}: Figure 2 presents the performance comparison on the large dataset using SSS@10 for effectiveness and latency for efficiency. SPLADE again demonstrates superior retrieval effectiveness compared to BM25. However, its latency is higher, reflecting its increased computational demands. Expanded-SPLADE provides a middle ground, improving effectiveness over BM25 while maintaining a more reasonable latency compared to SPLADE.

These results indicate that while SPLADE offers the best retrieval performance, Expanded-SPLADE provides a more balanced trade-off between effectiveness and efficiency across different dataset sizes.

\begin{figure}[h!]
    \centering
    \begin{tikzpicture}
        \begin{axis}[
            width=8cm, 
            height=6cm, 
            xlabel={\# FLOPS},
            ylabel={MRR@10},
            xmin=0, xmax=0.025,
            ymin=0.18, ymax=0.28,
            xtick={0.0, 0.005, 0.01, 0.015, 0.02},
            ytick={0.20, 0.22, 0.24, 0.26, 0.28},
            grid=major,
            legend pos=south east,
            legend style={font=\small},
            mark options={solid}
        ]
            \addplot[blue, thick, only marks, mark=square*, mark size=2.9pt] coordinates {
                (0.0027, 0.2030) 
            };
            \addlegendentry{BM25}
            \addplot[orange, thick, only marks, mark=square*, mark size=2.9pt] coordinates {
                (0.0100, 0.2549) 
            };
            \addlegendentry{Expanded-Splade}
            \addplot[red, thick, only marks, mark=square*, mark size=2.9pt] coordinates {
                (0.0217, 0.2733) 
            };
            \addlegendentry{Splade}
        \end{axis}
    \end{tikzpicture}
    \caption{Performance vs FLOPS on Small Dataset}
    \label{fig:distilsplade_flops}
\end{figure}
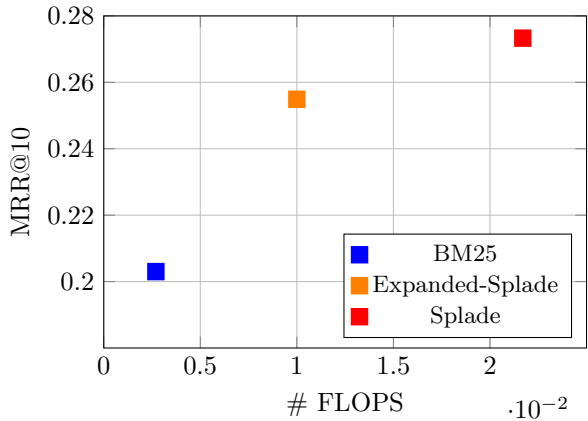

\begin{figure}[h!]
    \centering
    \begin{tikzpicture}
        \begin{axis}[
            width=8cm, 
            height=6cm, 
            xlabel={Latency},
            ylabel={SSS@10},
            xmin=0.5, xmax=2.0,
            ymin=0.65, ymax=0.8,
            xtick={0.5, 1.0, 1.5, 2.0},
            ytick={0.68, 0.71, 0.74, 0.77, 0.80},
            grid=major,
            legend pos=south east,
            legend style={font=\small},
            mark options={solid}
        ]
            \addplot[blue, thick, only marks, mark=square*, mark size=2.9pt] coordinates {
                (1.02, 0.6970) 
            };
            \addlegendentry{BM25}
            \addplot[orange, thick, only marks, mark=square*, mark size=2.9pt] coordinates {
                (1.20, 0.7642) 
            };
            \addlegendentry{Expanded-Splade}
            \addplot[red, thick, only marks, mark=square*, mark size=2.9pt] coordinates {
                (1.96, 0.7702) 
            };
            \addlegendentry{Splade}
        \end{axis}
    \end{tikzpicture}
    \caption{Performance vs Latency on Large Dataset}
    \label{fig:distilsplade_flops}
\end{figure}
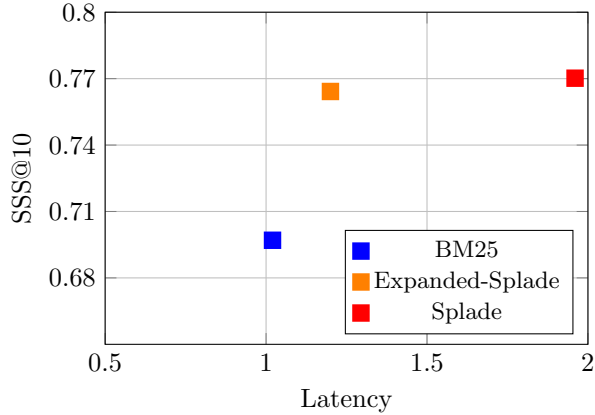

\subsection{Impact of Query Length on Latency}

\begin{table}[h]
    \centering
    \resizebox{\linewidth}{!}{%
        \begin{tabular}{lcc}
            \toprule
            \textbf{Model} & \textbf{Long Queries} & \textbf{Short Queries} \\
            \midrule
            BM25 & 1.02 (1$\times$) & 0.75 (1$\times$) \\
            SPLADE & 1.96 (1.92$\times$) & 4.75 (6.27$\times$) \\
            Expanded-SPLADE & 1.20 (1.27$\times$) & 2.39 (3.15$\times$) \\
            \bottomrule
        \end{tabular}
    }
    \caption{Latency Comparison Across Query Lengths on the Large Dataset}
    \label{tab:latency_comparison}
\end{table}

The previous results were based on long queries from the large dataset, where the average query length was 4.5 words. To further analyze the impact of query length on retrieval efficiency, we conducted additional experiments using short queries averaging 1.5 words, following \cite{thakur2023sprint}, which also explored retrieval latency variations in sparse retrieval models.

The results indicate that BM25 exhibits reduced latency for short queries (0.75 s) compared to long queries (1.02 s), suggesting that BM25 efficiently handles shorter inputs with minimal processing overhead. However, the trend is reversed for SPLADE-based models, where latency increases significantly for short queries. SPLADE, in particular, shows a notable increase from 1.96 s for long queries to 4.75 s for short queries, highlighting the computational burden introduced by its sparse expansion mechanism when handling shorter input texts. Expanded-SPLADE also experiences an increase in latency, rising from 1.20 s to 2.39 s, but the relative growth is smaller compared to SPLADE. These findings suggest that while sparse expansion methods benefit from longer query formulations, they encounter efficiency challenges when applied to extremely short queries, where their expansion processes introduce substantial computational overhead.

\subsection{Pruning}
Expanded-SPLADE exhibits significantly higher latency compared to BM25, with latency increasing more than threefold in some cases, as shown in Table \ref{tab:latency_comparison}. SPLADE shows even greater latency, exceeding six times that of BM25 for short queries. To mitigate this computational cost and improve retrieval efficiency, we conducted pruning experiments focusing on document-centric pruning, top-k query term selection, and Boolean Query with Term Threshold. By systematically adjusting pruning parameters, we aimed to strike a balance between retrieval effectiveness and efficiency.

\subsubsection{Top-k term selection}
To evaluate the impact of document-centric pruning and top-k query term selection on retrieval performance for large datasets, we conducted an experiment varying document k and query k values in the Expanded-SPLADE model. The results, presented in Table \ref{tab:pruning_effects}, highlight the trade-off between latency and retrieval effectiveness (SSS@10) on the large dataset.  

When applying only document-centric pruning, setting document k = 10 reduced latency from 1.20s to 0.91s, while SSS@10 improved slightly from 0.7642 to 0.7772. Increasing document k to 15 maintained this effectiveness with a marginal latency increase.  

When both document and query pruning were applied, reducing query k further decreased latency at the cost of effectiveness. For instance, with query k = 7, document k = 10, latency was minimized to 0.85s, while SSS@10 remained relatively high at 0.7725. However, further reducing query k to 5 led to a more noticeable drop in effectiveness (0.7655) despite additional latency reduction (0.82s).  

These results indicate that document-centric pruning alone can enhance both efficiency and effectiveness in large-scale retrieval scenarios, while query term pruning offers further latency gains but requires careful balancing to maintain retrieval performance.

\begin{table}[h]
    \centering
    \begin{tabular}{|c|c|c|c|}
        \hline
        \textbf{query k} & \textbf{document k} & \textbf{Latency} & \textbf{SSS@10} \\
        \hline
        \multicolumn{4}{|c|}{\textbf{Document}} \\
        \hline
        N/A  & N/A  & 1.20  & 0.7642 \\
        N/A  & 10   & 0.91  & 0.7772 \\
        N/A  & 15   & 0.94  & 0.7746 \\
        \hline
        \multicolumn{4}{|c|}{\textbf{Document + Query}} \\
        \hline
        7    & 10   & 0.85  & 0.7725 \\
        7    & 15   & 0.89  & 0.7706 \\
        5    & 10   & 0.82  & 0.7655 \\
        5    & 15   & 0.85  & 0.7627 \\
        \hline
    \end{tabular}
    \caption{Effect of document-centric pruning and top-k query term selection on latency and SSS@10 for the large dataset.}
    \label{tab:pruning_effects}
\end{table}

\subsubsection{Boolean Query with Term Threshold}

Figure~\ref{fig:boolean_query_results} presents the impact of different term threshold values on latency and retrieval performance (SSS@10) on the Large Dataset. The results indicate a clear trade-off between retrieval effectiveness and computational efficiency. At a 0\% threshold, the Expladed-SPLADE achieves the highest retrieval performance with SSS@10 = 0.7642, but incurs the highest latency of 1.2 seconds. As the term threshold increases, latency decreases significantly, with a reduction of nearly 85\% at an 80\% threshold, where latency drops to 0.18 seconds. However, this latency gain comes at the cost of retrieval performance, as SSS@10 declines steadily, reaching 0.2101 at the highest threshold. Notably, a threshold of 40\% provides a balanced trade-off, reducing latency by more than 60\% while maintaining over 94\% of the original retrieval performance. Similar trends are also observed for SPLADE. These findings suggest that moderate thresholding can substantially improve efficiency without severely degrading retrieval effectiveness, making it a viable strategy for optimizing sparse retrieval models in latency-sensitive applications.

\begin{figure}
    \centering
    \begin{adjustbox}{width=0.9\columnwidth}
        \begin{tikzpicture}
            \begin{axis}[
                xlabel={Latency},
                ylabel={SSS@10},
                grid=major,
                width=7cm,
                height=6cm,
                legend pos=south east,
                legend style={font=\small},
            ]
                \addplot[color=blue, mark=o, thick] coordinates {
                    (1.02, 0.6972)
                    (0.98, 0.6972)
                    (0.63, 0.6947)
                    (0.27, 0.6484)
                    (0.10, 0.4008)
                };
                \addlegendentry{BM25}
                \addplot[color=orange, mark=o, thick] coordinates {
                    (1.2, 0.7642)
                    (0.88, 0.7614)
                    (0.46, 0.7209)
                    (0.22, 0.5205)
                    (0.18, 0.2101)
                };
                \addlegendentry{Expanded-Splade}
                \addplot[color=red, mark=o, thick] coordinates {
                    (1.96, 0.7702)
                    (1.06, 0.7717)
                    (0.64, 0.7439)
                    (0.43, 0.5258)
                    (0.36, 0.1756)
                };
                \addlegendentry{Splade}
            \end{axis}
        \end{tikzpicture}
    \end{adjustbox}
    \caption{Experimental Results of Boolean Query with Term Threshold on the Large Dataset using Expanded-SPLADE. The term threshold increases in 20\% increments from right to left, starting from 0\% on the rightmost node and progressing to 80\% on the leftmost node.}
 
    \label{fig:boolean_query_results}
\end{figure}
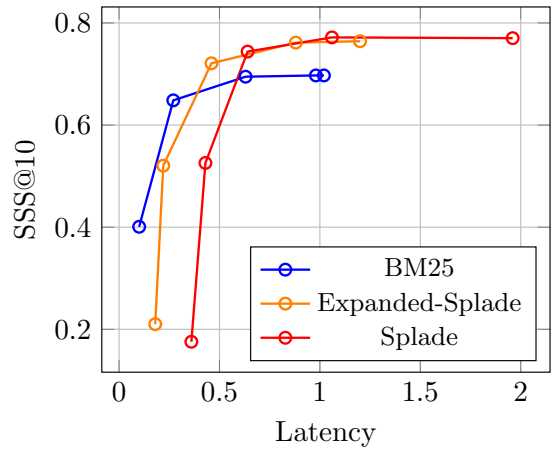

\begin{figure*}
    \centering
    \begin{adjustbox}{width=\columnwidth*2} 
        \begin{tikzpicture}
            \begin{axis}[
                xlabel={Latency},
                ylabel={SSS@10},
                grid=major,
                xmin=0.0, xmax=2.0,
                ymin=0.0, ymax=0.9,
                width=14cm, 
                height=6cm,
                legend pos=south east,
                legend style={font=\tiny},
            ]
                \addplot[color=blue, mark=*, only marks] coordinates {
                    (1.02, 0.6972)
                    (0.98, 0.6972)
                    (0.63, 0.6947)
                    (0.27, 0.6484)
                    (0.10, 0.4008)
                };
                \addlegendentry{BM25}
                \addplot[color=green, mark=*, only marks] coordinates {
                    (1.2, 0.7642)
                    (0.88, 0.7614)
                    (0.46, 0.7209)
                    (0.22, 0.5205)
                    (0.18, 0.2101)
                };
                \addlegendentry{ExpandedSplade-qk0-dk0}
                \addplot[color=purple, mark=triangle*, only marks] coordinates {
                    (0.91, 0.7772)
                    (0.58, 0.7422)
                    (0.14, 0.5686)
                    (0.08, 0.2622)
                    (0.07, 0.0780)
                };
                \addlegendentry{ExpandedSplade-qk0-dk10}
                \addplot[color=brown, mark=pentagon*, only marks] coordinates {
                    (0.82, 0.7655)
                    (0.82, 0.7654)
                    (0.71, 0.7653)
                    (0.21, 0.7617)
                    (0.04, 0.5601)
                };
                \addlegendentry{ExpandedSplade-qk5-dk10}
                \addplot[color=gray, mark=square*, only marks] coordinates {
                    (0.85, 0.7725)
                    (0.75, 0.7726)
                    (0.47, 0.7730)
                    (0.05, 0.5427)
                    (0.04, 0.2426)
                };
                \addlegendentry{ExpandedSplade-qk7-dk10}
                \addplot[color=red, mark=*, only marks] coordinates {
                    (1.96, 0.7702)
                    (1.06, 0.7717)
                    (0.64, 0.7439)
                    (0.43, 0.5258)
                    (0.36, 0.1756)
                };
                \addlegendentry{Splade-qk0-dk0}
                \addplot[color=pink, mark=triangle*, only marks] coordinates {
                    (1.07, 0.7877)
                    (0.54, 0.7343)
                    (0.17, 0.4041)
                    (0.12, 0.1363)
                    (0.10, 0.0323)
                };
                \addlegendentry{Splade-qk0-dk10}
                \addplot[color=orange, mark=pentagon*, only marks] coordinates {
                    (0.89, 0.7490)
                    (0.91, 0.7490)
                    (0.77, 0.7489)
                    (0.23, 0.7432)
                    (0.04, 0.5548)
                };
                \addlegendentry{Splade-qk5-dk10}
                \addplot[color=yellow, mark=star, only marks] coordinates {
                    (0.94, 0.7733)
                    (0.83, 0.7733)
                    (0.54, 0.7722)
                    (0.06, 0.5511)
                    (0.05, 0.2593)
                };
                \addlegendentry{Splade-qk7-dk10}
            \end{axis}
        \end{tikzpicture}
    \end{adjustbox}
    \caption{Effect of Combining Document Pruning, Query Term Selection, and Boolean Query Thresholding on Retrieval Efficiency and Performance in the Large Dataset. For each model configuration (e.g., Splade-qk0-dk10), the term threshold increases in 20\% increments from 0\% (rightmost node) to 80\% (leftmost node). Here, qk represents the k value for query term selection, and dk represents the k value for document-centric pruning. A value of k=0 indicates that no pruning was applied for the corresponding method.}
    \label{fig:combining_results}
\end{figure*}
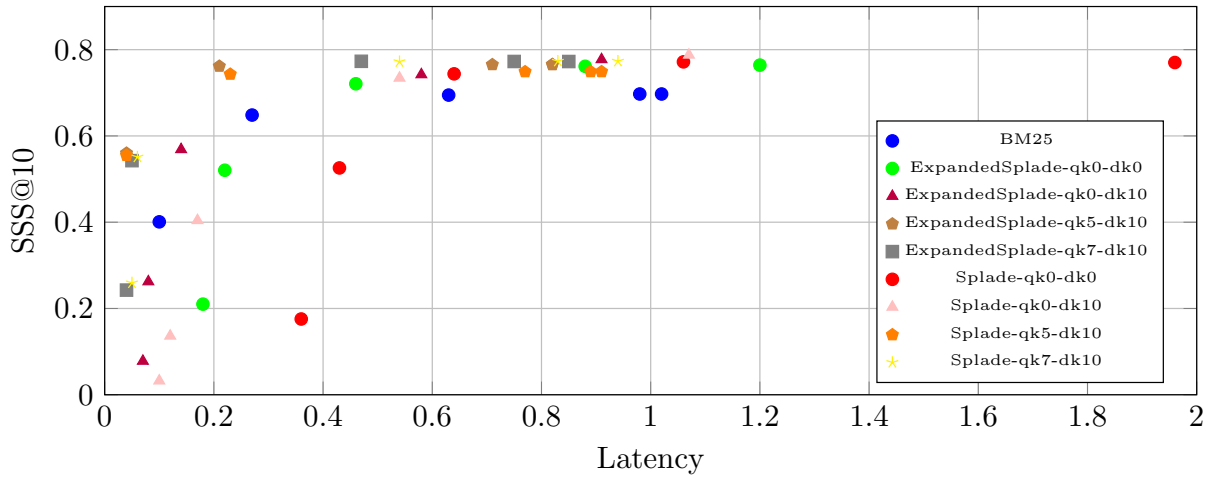

\subsubsection{Evaluating the Combined Effect of Pruning}
The final experiment investigates the combined effect of document-centric pruning, top-k query term selection, and boolean query with term threshold on retrieval efficiency and effectiveness. As shown in Figure~\ref{fig:combining_results}, applying these techniques in combination leads to significant latency reductions while maintaining competitive retrieval performance across all models. While each method individually contributes to efficiency gains, their combined application further amplifies these benefits. Increasing the term threshold consistently reduces latency, with the trade-off varying based on the degree of document pruning and query term selection. SPLADE and Expanded-SPLADE without pruning (qk0-dk0) exhibit the highest latencies, but introducing document pruning (qk0-dk10) and query term selection (qk5-dk10) leads to substantial efficiency improvements with only moderate performance degradation. Notably, ExpandedSplade-qk7-dk10 demonstrates the most balanced trade-off, achieving latency reductions close to BM25 while maintaining a high SSS@10 score. These findings highlight that carefully tuning the combination of pruning, query term selection, and thresholding strategies can significantly optimize retrieval efficiency while mitigating the impact on effectiveness.

\section{Conclusion}

In this paper, we have compared the retrieval performance and computational efficiency of BM25, SPLADE, and Expanded-SPLADE on a billion-scale dataset. Our experimental results highlight the strengths and weaknesses of each model in terms of both effectiveness and efficiency. While SPLADE offers significant improvements over BM25 in retrieval effectiveness, it comes at the cost of increased computational resources. Expanded-SPLADE provides a more balanced trade-off, achieving improved effectiveness with reduced computational overhead compared to SPLADE.

Moreover, we explored pruning strategies, such as document-centric pruning, top-k query term selection and boolean query with term threshold, to further optimize the efficiency of these models. These strategies successfully reduced latency while maintaining high retrieval quality, making SPLADE models a viable solution for real-time web search applications.

In conclusion, sparse retrieval models like SPLADE and Expanded-SPLADE show great potential for large-scale information retrieval tasks, but careful consideration of computational costs and the use of pruning techniques is essential for their practical deployment.

\end{document}